\documentstyle[epsf,12pt]{article}

\newcommand{\reseteqnum}{\setcounter{equation}{0}}

\renewcommand{\slash}[2]{#1\kern-#2pt\mbox{\it/}}
\newcommand{\slp}{p\kern-6.0pt\mbox{\it/}}

\newcommand{\simle}{{\;}_{\displaystyle_{\displaystyle\sim}}\kern-14.9pt<}
\newcommand{\simge}{{\;}_{\displaystyle_{\displaystyle\sim}}\kern-14.9pt>}

\newcommand{\xsize}{10cm}

\catcode`\@=11
\newbox\tempboxa
\newdimen\captionboxsubcount
\def\capsize#1{\captionboxsubcount=#1pt}
\newdimen\captionboxsub
\captionboxsub=\hsize \advance\captionboxsub by -\captionboxsubcount
\advance\captionboxsub by -\captionboxsubcount
\long\def\@makecaption#1#2{
 \setbox\@tempboxa\hbox{#1: #2}
 \ifdim \wd\@tempboxa >\captionboxsub
\rightskip=\captionboxsubcount \leftskip=\captionboxsubcount #1: #2
\else \hbox to\hsize{\hfil\box\@tempboxa\hfil}
 \fi}
\catcode`\@=12
\capsize{30}

\begin{document}

\begin{titlepage}
\begin{flushright}
KUNS-1373 \\
HE(TH) 95/23 \\
hep-ph/9512375 \\
\end{flushright}

\begin{center} \Large
     CHIRAL SYMMETRY RESTORATION \\
     AT FINITE TEMPERATURE AND CHEMICAL POTENTIAL \\
     IN THE IMPROVED LADDER APPROXIMATION
\end{center}
\bigskip

\begin{center} \Large
        Yusuke Taniguchi
\footnote{e-mail address :
{\tt tanigchi@gauge.scphys.kyoto-u.ac.jp}}
and
        Yuhsuke Yoshida
\footnote{e-mail address :
{\tt yoshida@gauge.scphys.kyoto-u.ac.jp}}
\end{center}
\bigskip

\begin{center} \large \it
         Department of Physics, Kyoto University \\
         Kyoto 606-01, Japan
\end{center}

\begin{center} \Large \bf
Abstract
\end{center}

\begin{quote}
The chiral symmetry of QCD is studied at finite temperature and
chemical potential using the Schwinger-Dyson equation in the improved
ladder approximation.
We calculate three order parameters; the vacuum expectation value of
the quark bilinear operator, the pion decay constant and the quark
mass gap.
We have a second order phase transition at the temperature $T_c=169$
MeV along the zero chemical potential line, and a first order phase
transition at the chemical potential $\mu_c=598$ MeV along the zero
temperature line.
We also calculate the critical exponents of the three order
parameters.
\end{quote}

\vspace{20pt}
\noindent
PACS numbers 11.10.Wx, 11.15.Tk, 11.30.Qc, 11.30.Rd, 12.38.-t\,.

\end{titlepage}

\section{Introduction}
\label{sect:intro}
\reseteqnum

The chiral symmetry in QCD is dynamically broken at zero temperature.
This feature is confirmed by the fact that the pion is the
Nambu-Goldstone boson accompanied with this symmetry breaking.
On the other hand, it is shown that spontaneously broken symmetries
restore at sufficiently high temperature (and/or chemical potential)
in some simple models.\cite{SymRes}
Then, the same restoration is also expected to hold for the
dynamically broken chiral symmetry in QCD.
This phenomena is widely believed to be seen in heavy-ion collisions,
the early universe and the neutron stars.

There are various attempts to study the phase diagram and critical
behavior.
In order to do them we need any non-perturbative treatments
such as $\varepsilon$ or $1/N$ expansion, lattice simulation, the
Schwinger-Dyson equation and so on.
Based on universality arguments it is expected that critical
phenomena of finite temperature QCD is described by a three
dimensional linear $\sigma$ model with the same global
symmetry, in which the $\varepsilon$ expansion is used.\cite{univ}
Lattice simulations are powerful tools to study the QCD at finite
temperature.\cite{KSlattice,mulattice,lattice,KEKlattice}
The Schwinger-Dyson equation in the improved ladder approximation is
solved with further approximations and give the dynamical symmetry
restoration.\cite{BCS,Kocic,Akiba,ER,BCCGP}
Nambu-Jona-Lasinio models, as phenomenological models of QCD, provide
us with useful pictures about the dynamical chiral symmetry breaking
and its restoration.\cite{HatsuKuni,AY,LKW}
In these three approaches it is indicated that there is a second order
phase transition at $(T,\mu)=(T_c,0)$ and a first order one at
$(T,\mu)=(0,\mu_c)$ in the case of the two massless flavors.
The phase transition points are found to be of order $T_c\sim 200$,
$\mu_c\sim 400$ MeV.

In this paper we use the Schwinger-Dyson equation in the improved
ladder approximation.
The advantages of this approach are that it is a convenient tool to
study the nature of the chiral symmetry, and we easily introduce
fermions coupling to the gluon in the chiral limit at finite
temperature and chemical potential.
Further, we have no degrees of freedom (parameters) to fit the
physical observables, and we obtain a definite answer.

In previous attempts further approximations are introduced in
addition to the ladder approximation.
Some non-perturbative approximation can violate the chiral symmetry,
and reliable results cannot be obtained.
We would like to obtain the results keeping the chiral symmetry within
the framework of the Schwinger-Dyson equation only in the improved
ladder approximation.

We calculate the values of the three order parameters; the quark mass
gap, the vacuum expectation value of the quark bilinear operator
$\langle\overline\psi\psi\rangle$ and the pion decay constant.
We have a second order phase transition at $T_c=169$ MeV along the
$\mu=0$ line and a first order phase transition at $\mu_c=598$ MeV
along the $T=0$ line.
The critical exponents of the above three order parameters are
extracted at $(T,\mu)=(T_c,0)$, which shows that our formulation is
different from mean field theories.

This paper is organized as follows.
In section~\ref{sect:SD}, we show basic ingredients to study the chiral
symmetry restoration using the Schwinger-Dyson equation in the
improved ladder approximation.
The expressions of the three order parameters are given in terms of
the quark mass function.
The Pagels-Stokar formula is used to calculate the pion decay
constant.
In section~\ref{seq:results} we give our numerical results.
The Schwinger-Dyson equation is numerically solved using a iteration
method.
We determine the positions and the orders of phase transitions.
We also extract the critical exponents.
Summary and discussion are found in section~\ref{sect:sam-dis}.

\section{Schwinger-Dyson Equation at Finite Temperature and Chemical
  Potential}
\label{sect:SD}
\reseteqnum

The restoration of spontaneously broken symmetry occurs at finite
temperature and chemical potential.\cite{SymRes}
The phenomena is described in terms of the imaginary time formalism in
gauge theories.\cite{ITF,PIF}

In this section we show basic ingredients to solve the Schwinger-Dyson
equation at finite temperature and chemical potential.
Then, we study the dynamical chiral symmetry breaking and its
restoration.
In this paper all dimensionful quantities are rescaled by using the
$\Lambda_{QCD}$, otherwise stated.

We consider the QCD with massless $u$ and $d$ quarks, and then there
is a $SU(2)_L\times SU(2)_R$ chiral symmetry.
There are several probes to investigate the chiral symmetry such as
the quark mass gap, the vacuum expectation value of the quark
bilinear operator (VEV) and the decay constant of the pion.
Those are evaluated in terms of the quark mass function $\Sigma(p)$.
The mass function is determined by the Schwinger-Dyson equation.
We use the improved ladder approximation to solve this equation.
We work with three flavor $\beta$-function for the running coupling,
since the $s$ quark also contributes to the running of the coupling in
the concerned energy range.

At zero temperature this approximation provides us with a convincing
result of the dynamical chiral symmetry breaking and good values of
the lowest-lying meson masses.
So, we expect that the approximation gives good result even in the
case of finite temperature.

\begin{figure}[htbp]
\epsfxsize=10cm
\begin{center}
\ \epsfbox{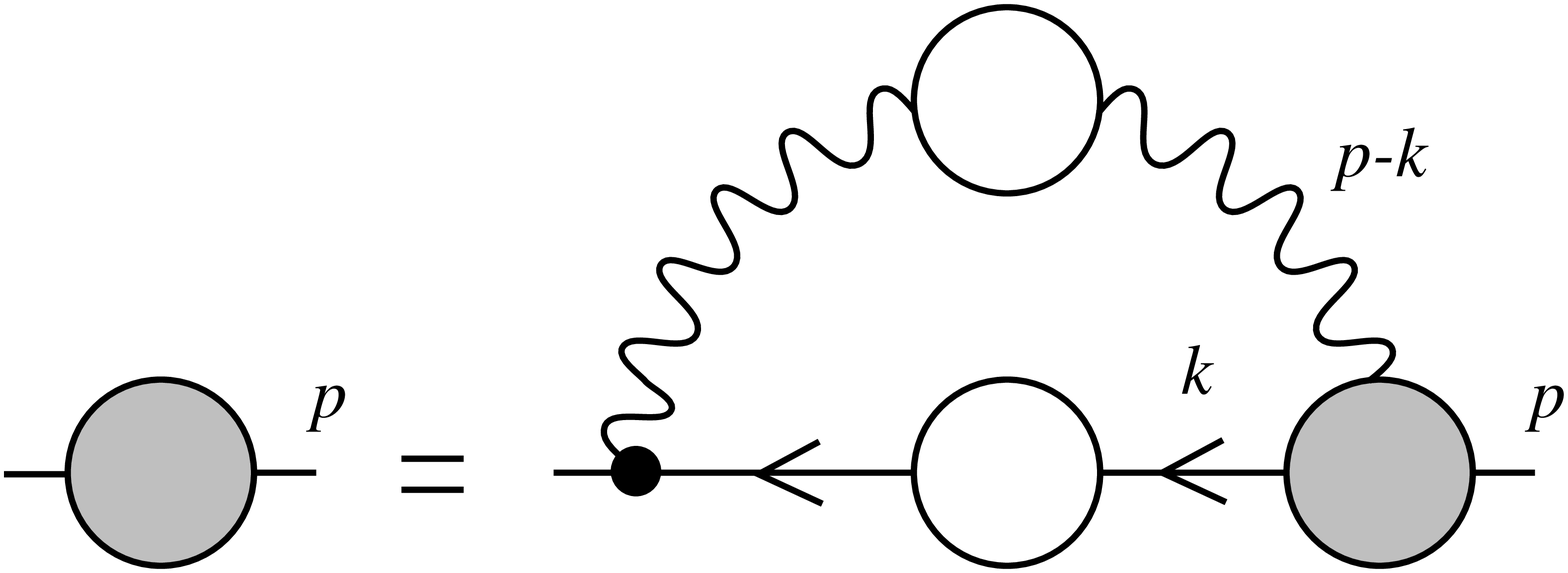}
\vspace{-5pt}
\caption[]{
The Feynman diagram of the Schwinger-Dyson equation.
We have the same diagram for the equation at finite temperature and
chemical potential.
}
\label{fig:SDeq}
\end{center}
\end{figure}

To take the effect of finite temperature into account, we work in
the imaginary time formalism\cite{ITF,PIF}.
Let us start with writing down the Schwinger-Dyson equation as in
Fig.~\ref{fig:SDeq}.
The diagram is exactly the same as that in the zero temperature QCD,
since the difference between the usual (zero temperature) field theory
and finite temperature field theory stems from the boundary effect of
time only.
The time components of quark and gluon momenta become discrete.
Since quark fields have anti-periodic boundary condition in the
imaginary time direction, we have
\begin{eqnarray}
\label{p0}
p^0 &=& 2\pi iT\left(n+\frac{1}{2}\right) ~,\nonumber\\
k^0 &=& 2\pi iT\left(m+\frac{1}{2}\right) ~,
\end{eqnarray}
where $n,m \in \mbox{\boldmath$Z$}$.
When the chemical potential $\mu$ is introduced, the time component of
$p$ in the quark propagator $S_F(p)$ is modified as
\begin{equation}
  \label{p0mu}
  p_0 \to p_0 - \mu ~.
\end{equation}
The momentum integration is modified to the summation:
\begin{equation}
\label{sum}
\int \frac{d^4k}{(2\pi)^4i}
 \quad \rightarrow \quad
T \sum_{m=-\infty}^\infty\int\frac{d^3k}{(2\pi)^3} ~.
\end{equation}
Then, modifying the Schwinger-Dyson equation at zero temperature
according to Eqs.~(\ref{p0}), (\ref{p0mu}) and (\ref{sum}), the
Schwinger-Dyson equation at finite temperature is given as
\begin{equation}
\slp - S_F(p)^{-1} = T \sum_{m=-\infty}^{\infty}
\int\frac{d^3k}{(2\pi)^3} C_2g^2(p,k)
K_{\mu\nu}(p-k)\gamma^\mu S_F(k) \gamma^\nu ~,\label{SDeq}
\end{equation}
where $C_2$ is the second Casimir invariant and $-K_{\mu\nu}$ is the
gluon tree propagator in the Landau gauge:
\begin{equation}
K_{\mu\nu}(l) = \frac{1}{-l^2}
\left(g_{\mu\nu}-\frac{l_\mu l_\nu}{l^2}\right) ~.
\end{equation}
The quantity $g^2(p,k)$ is a one-loop running coupling depending on
the momenta $p$ and $k$.
In order to describe a property of QCD we use the following form for
the running coupling:\cite{ABKMN}
\begin{equation}
\label{g2}
g^2(p,k) =
   \frac{1}{\beta_0} \times \left\{\begin{array}{ll}
\displaystyle \frac{1}{t} & \mbox{ if $t_F < t$ } \smallskip\\
\displaystyle \frac{1}{t_F} + \frac{(t_F - t_C)^2
   - (t - t_C)^2}{2t_F^2(t_F - t_C)} &\smallskip
   \mbox{ if $ t_C < t < t_F$ } \\
\displaystyle \frac{1}{t_F} + \frac{(t_F - t_C)}{2t_F^2} &
   \mbox{ if $ t < t_C$ } \smallskip
   \end{array}\right.~,
\end{equation}
where $t = \ln (p^2+k^2)$, $t_C\equiv-2.0$,
$\beta_0 = (11N_c-2N_f)/(48\pi^2)$ is the coefficient of one-loop
$\beta$-function and $t_F$ is a parameter needed to regularize the
divergence of the running coupling at the QCD scale $\Lambda_{QCD}$.
We call $t_F$ the infrared regularization parameter.
$N_c=3$ and $N_f=3$ are the number of colors and flavors,
respectively.
The running coupling $g^2$ is smoothly interpolated between the
ordinary one-loop running coupling form at $t>t_F$ and a constant
value in the low energy region.
As will be shown, the results do not depend on this particular
infrared regularization.
By virtue of the running effect of the coupling the resultant mass
function, which is determined by Eq.~(\ref{SDeq}), reproduces the
exact behavior in the high energy region.\cite{Politzer}
Notice that this property is needed to preserve the chiral
symmetry.\cite{HigaMira}

The quark propagator is expanded by three $SO(3)$ invariant amplitudes
as
\begin{equation}
S_F(p) = \frac{1}{\Sigma(p) + (\mu+B(p))\gamma^0 - A(p)\slp }
{}~.\label{SF}
\end{equation}

At the vanishing temperature and chemical potential limits ($T,
\mu\rightarrow0$) the choice of Landau gauge allows us to obtain
\begin{eqnarray}
A(p) &=& 1 ~,\nonumber\\
B(p) &=& 0 ~.\label{AB}
\end{eqnarray}
Although we are studying in finite temperature and chemical potential,
we assume the relations (\ref{AB}) for simplicity.
We expect that the relation (\ref{AB}) is not changed so much in the
case of low temperature and chemical potential.
As shown later, the phase transition line of the chiral symmetry
restoration lies in that region; $T_c,\mu_c \simle \Lambda_{QCD}$.
Then, the result should not change qualitatively.

Now, substituting Eq.~(\ref{SF}) into Eq.~(\ref{SDeq}) under the
condition (\ref{AB}), we obtain
\begin{equation}
\Sigma_n(x) = \sum_{m=-\infty}^{\infty}\int ydy
K_{nm}(x,y) \frac{\Sigma_m(y)}
{\left(2\pi T(m+\frac{1}{2})+i\mu\right)^2+y^2+\Sigma_m(y)^2} ~,
\label{SDcompo}
\end{equation}
where $x = |\mbox{\boldmath$p$}|$ and $y = |\mbox{\boldmath$k$}|$ and
\begin{equation}
K_{nm}(x,y) = \frac{3TC_2g^2(p,k)}{8\pi^2xy} \;
\ln\left(\frac{4\pi^2T^2(n-m)^2+(x+y)^2}
              {4\pi^2T^2(n-m)^2+(x-y)^2}\right) ~.\label{kernel}
\end{equation}
Notice that we have the $SO(3)$ rotational invariance but not
$SO(3,1)$.
The mass function $\Sigma(p)$ is a function of $p^0$ and
$\mbox{\boldmath$p$}^2$, and is rewritten as $\Sigma_n(x)$.
In the presence of the chemical potential we easily find from
Eq.~(\ref{SDcompo}) that the mass function takes a complex value
satisfying the relation
\begin{equation}
\label{complex}
\Sigma_{-n}(x)^* = \Sigma_{n-1}(x)~~~\mbox{for}~~~n=1,2,\cdots~.
\end{equation}

We solve the Schwinger-Dyson equation (\ref{SDcompo}) numerically by
an iteration (relaxation) method.
The momentum valuables $x$, $y$ are discretized to be
\begin{equation}
\label{x}
x \quad \rightarrow \quad x_n
= \exp\left(\Lambda_{IR}+(\Lambda_{UV}-\Lambda_{IR})
\frac{n-1}{N_{SD}-1} \right)~,
\end{equation}
where $n = 1, 2, \cdots, N_{SD}$ and similarly for $y$.
We divide $\ln x$ and $\ln y$ into $N_{SD}$ points.
The quantity $\Lambda \equiv \exp\Lambda_{UV}$ defines the ultraviolet
cutoff for the space component of momenta.
Therefore an $SO(3)$ symmetric cutoff $\Lambda$ is introduced, which
is needed for numerical calculation.
The momentum region of the time component is properly truncated
so that the support of the mass function is covered well, as well as
that of the space component.
We have integrable singularities at $(n,x) = (m,y)$ stemming from the
tree level gluon pole, but these should be regularized in the
numerical calculation.
In order to avoid this singularity we apply a two-point splitting
prescription
\begin{equation}
K_{nm}(x,y) \quad\rightarrow\quad
 \frac{1}{2}\Big(K_{nm}(x,y_+)+K_{nm}(x,y_-)\Big) ~,
\end{equation}
with $y_\pm = y\exp(\pm (\Lambda_{UV}-\Lambda_{IR})/(4N_{SD}))$.
The validity of this prescription is checked by using the conventional
(zero temperature) Schwinger-Dyson equation.

After obtaining the mass function, we immediately evaluate the pion
decay constant by using the Pagels-Stokar formula at finite
temperature:
\begin{equation}
f_\pi(T,\mu)^2 = \frac{2N_cT}{\pi^2}\sum_n \int_0^\infty x^2dx\,
\frac{\Sigma_n(x)\bigg(\Sigma_n(x)-
 \displaystyle\frac{x}{3}\frac{d\Sigma_n(x)}{dx}\bigg)}
{\left(\Big(2\pi T(n+\frac{1}{2})+i\mu\Big)^2+
 x^2+\Sigma_n(x)^2\right)^2} ~.
\label{PS-formula}
\end{equation}
We also calculate the VEV
\begin{equation}
\langle\overline uu\rangle_\Lambda =
\langle\overline dd\rangle_\Lambda =
\frac{2N_cT}{\pi^2}\sum_n \int_0^\Lambda x^2dx\,
\frac{\Sigma_n(x)}{\Big(2\pi T(n+\frac{1}{2})+i\mu\Big)^2+
 x^2+\Sigma_n(x)^2} ~.
\end{equation}
The VEV is renormalized at 1 GeV via
\begin{equation}
\langle\overline\psi\psi\rangle_{1\rm GeV}
=
\left(\frac{\ln\big(1\rm GeV\big)}{\ln\Lambda}\right)^\frac
{\scriptstyle 11N_c-2N_f}{\scriptstyle 9C_2}
\langle\overline\psi\psi\rangle_\Lambda ~,
\end{equation}
where $\psi = u,d$.

\section{Numerical Results}
\label{seq:results}
\reseteqnum

\begin{figure}[htbp]
\epsfxsize=7cm
\begin{center}
\ \epsfbox{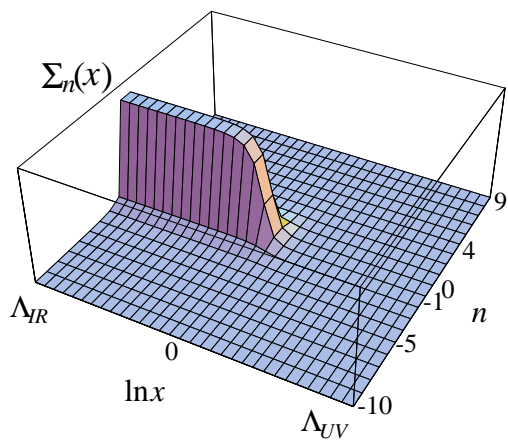}
\vspace{-5pt}
\caption[]{
A typical form of the mass function.
We put $T=90$, $\mu=0$ MeV.
The integer $n$ specifies the time component of the momentum as in
Eq.~(\ref{p0}) and $x$ is given in Eq.~(\ref{x}).
}
\label{fig:mf}
\end{center}
\end{figure}

In this section we solve the Schwinger-Dyson equation numerically by
an iteration method.
We start with an initial form of the mass function and input it in the
R.H.S. of Eq.(\ref{SDcompo}).
Performing the integration of $x$ and the summation of $n$, we have
an updated form of the mass function, which is taken as a more
suitable trial form.
After sufficient iterations the functional form converges giving the
true solution with enough accuracy.
The convergence of the solution is very rapid off the phase transition
regions.
A typical form of the mass function is shown in Fig.~\ref{fig:mf} at
$T=90$, $\mu=0$ MeV.
The mass function dumps so fast in the $p_4 \equiv -ip_0$ direction
that the values of the mass function at $n \simge 4$ are much smaller
than that at $n=0$ ($\Sigma_n(x) \sim 10^{-2} \times \Sigma_0(x)$).
This implies a dimensional reduction at sufficiently high
temperature; i.e., four dimensional theory at finite temperature
belongs to the same universality class as that of a three dimensional
one with the same symmetry.

We use the VEV and the pion decay constant as order parameters of the
chiral symmetry.
The dynamical mass of the quark itself is also an order parameter.

We determine the value of $\Lambda_{QCD}$ from the
experimental result $f_\pi = 93$ MeV at $T=\mu=0$,
and we have $\Lambda_{QCD} = 592$ MeV using Eq.~(\ref{PS-formula})
at $t_F=0.5$.
This is our result of $\Lambda_{QCD}$ obtained by using the one-loop
$\beta$-function.
In this paper we put the infrared regularization parameter $t_F=0.5$
and show later that the physical observables as well as the phase
transition line do not depend on $t_F$.

\subsection{Zero chemical potential case}
\begin{figure}[htbp]
\epsfxsize=\xsize
\begin{center}
\ \epsfbox{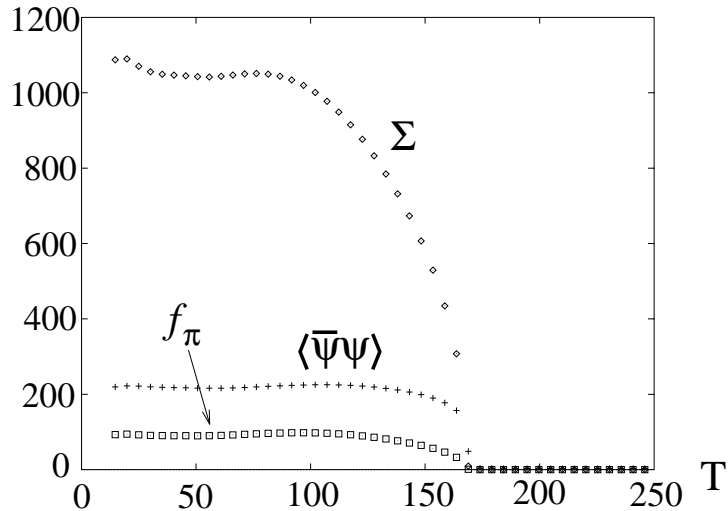}
\vspace{-5pt}
\caption[]{
The functional forms of the order parameters (the mass function,
the VEV and the pion decay constant) for $T$ along the $\mu=0$ line.
}
\label{fig:T}
\end{center}
\end{figure}

First, we study the phase transition along the $\mu=0$ line.
The phase transition point is defined so that three order parameters
of the mass gap $\Sigma_{n=0}(x=0)$, the
VEV $\langle\overline\psi\psi\rangle$ and the pion decay constant
$f_\pi$ vanish.
We show the temperature dependences of these order parameters in
Fig.~\ref{fig:T} with two massless flavors.
The $SU(2)_L \times SU(2)_R$ chiral symmetry restores at
$T=T_c=169$ MeV.
We have a second order phase transition.
We have the same result as that of lattice
simulations\cite{lattice,KEKlattice} with two flavors in which the
phase transition is second order at $T_c \sim 200$ MeV.

The ladder approximation, used here, gives no flavor dependence, since
the dependence essentially comes through only the running effect of
the coupling.
Whereas the flavor dependence is suggested by the universality
arguments\cite{univ} and it is confirmed by lattice
simulations\cite{KSlattice,lattice,KEKlattice}, where we have a second
order phase transition at $N_f=2$ and first order ones at $N_f\ge 3$.
The Nambu-Jona-Lasinio models\cite{HatsuKuni,AY,LKW} imply that the
inclusion of the effect of $U(1)_A$ anomaly, so called the instanton
effect, allows us to obtain the same flavor dependence as that in
the lattice simulations.

The important point in studying the chiral symmetry is that the
approximation used should preserve the symmetry.
Fortunately, the ladder approximation itself is consistent with the
chiral symmetry.\cite{HigaMira}
However, many investigations violate the chiral symmetry, where further
approximations are used in addition to the ladder\cite{BCS,Akiba,ER}.
In order to reserve the symmetry the high energy behavior of the quark
mass function must be consistent with the result of the operator
product expansion (OPE):\cite{HigaMira}
\begin{equation}
\label{OPE}
\Sigma_n (\mbox{\boldmath $p$}^2) \sim \frac{g^2(x)}{x}
\left(\ln x \right)^{\textstyle\frac{9C_2}{11N_c-2N_f}}
{}~~~\mbox{as}~~~ x\equiv p_4^2+\mbox{\boldmath $p$}^2 \sim \infty~.
\end{equation}
Here we should notice that even in the finite temperature case the
high energy behavior of the mass function is the same as that of zero
temperature case, since the temperature effect is suppressed in the
high energy region.
In Refs.~\cite{BCS,Akiba,ER} their further approximation
($\Sigma={\rm const.}$) does not satisfy Eq.~(\ref{OPE}), on the other
hand in Ref.~\cite{BCCGP} they adopt an ansatz consistent with
Eq.~(\ref{OPE}) up to the logarithmic correction.
While our formalism exactly reproduce the OPE result (\ref{OPE}).

\begin{figure}[htbp]
\begin{center}
\epsfxsize=\xsize
\ \epsfbox{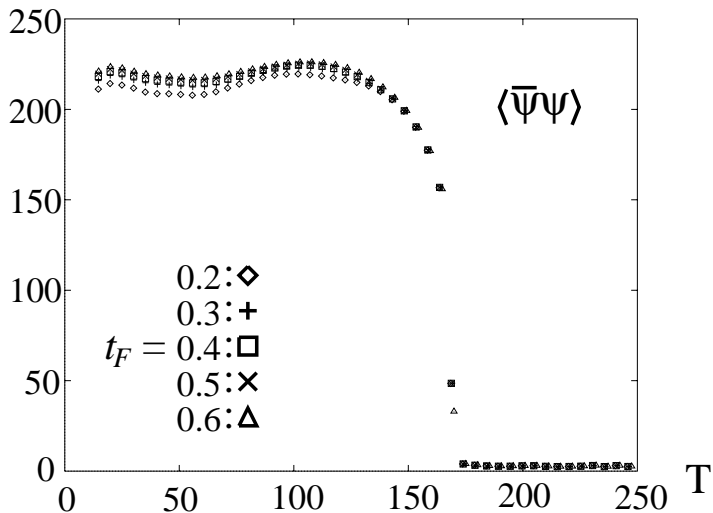}
\vspace{-5pt}
\caption[]{
The $t_F$ dependence of the VEV.
It changes by $5\%$ at worst against $t_F$.
}
\label{fig:tF_vev}
\end{center}
\end{figure}

\begin{figure}[htbp]
\begin{center}
\epsfxsize=\xsize
\ \epsfbox{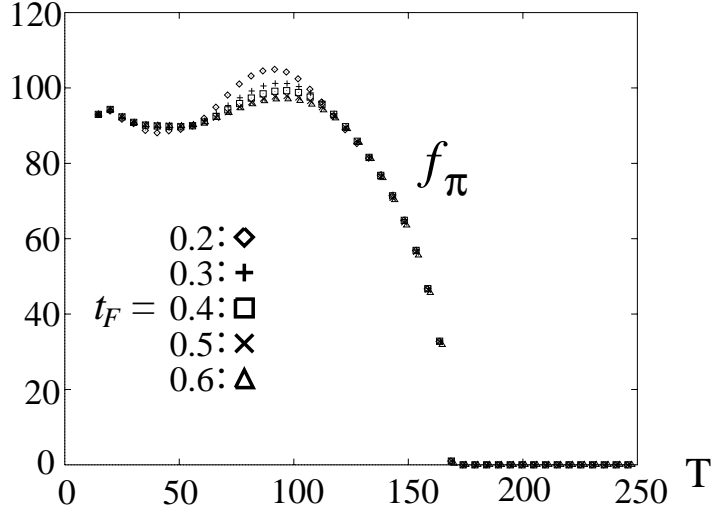}
\vspace{-5pt}
\caption[]{
The $t_F$ dependence of the pion decay constant.
It changes by $8\%$ at worst against $t_F$.
}
\label{fig:tF_fpi}
\end{center}
\end{figure}

We check the dependence of the order parameters on the infrared
regularization parameter $t_F$.
The physical observables, $\langle\overline\psi\psi\rangle_{1\rm GeV}$
and $f_\pi(T)$, should not depend on the parameter $t_F$.
We confirm this requirement.
The dependence of the VEV and the $f_\pi(T)$ are shown in
Figs.~\ref{fig:tF_vev} and \ref{fig:tF_fpi}.
The values of the VEV and the $f_\pi$ change, at worst, by $5\%$ and
$8\%$ against $t_F = 0.2 \sim 0.6$, respectively.
These values of $t_F=0.2\sim0.6$ correspond to those of the running
coupling $g^2(p\!=\!k\!=\!0)=570\sim92.6$.
Moreover, the phase transition point is fairly stable against $t_F$.
\begin{figure}[htbp]
\begin{center}
\epsfxsize=\xsize
\ \epsfbox{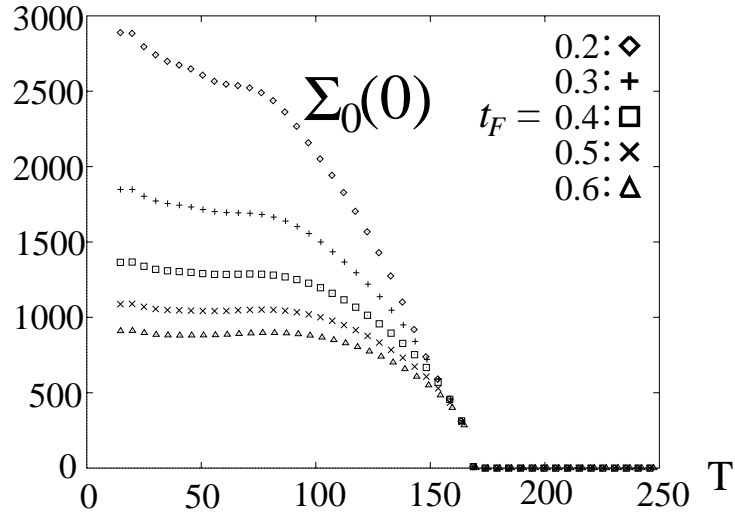}
\vspace{-5pt}
\caption[]{
The $t_F$ dependence of the mass function.
We have a strong dependence.
}
\label{fig:tF_mf}
\end{center}
\end{figure}
We conclude that there is no $t_F$ dependence of the physical
observables and the position of the phase transition point.

We also show the dependence of the mass function in
Fig.~\ref{fig:tF_mf}.
We have a strong $t_F$ dependence.
Since the mass function is not a physical observable it may depend on
the regularization parameter $t_F$.
However the phase transition point determined using the mass function
is fixed.

\begin{figure}[htbp]
\begin{center}
\epsfxsize=\xsize
\ \epsfbox{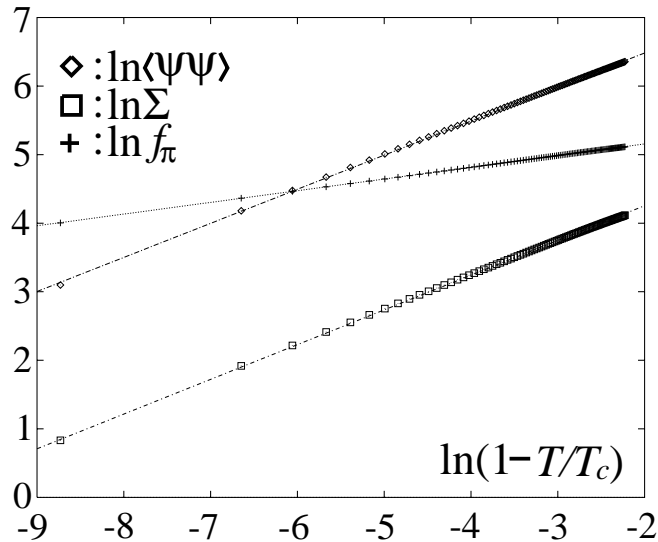}
\vspace{-5pt}
\caption[]{
The $\chi^2$ fittings for extracting the critical exponents of the
VEV, the mass gap and the pion decay constant.
We draw the best fitted lines of the form $A+\gamma\ln(1-T/T_c)$.
}
\label{fig:exponents}
\end{center}
\end{figure}

Let us examine the critical behavior of the system.
Since we have a second order phase transition at $(T,\mu) = (T_c,0)$,
the three order parameters behave near the phase transition point as
\begin{eqnarray}
\langle\overline\psi\psi\rangle_{1\rm GeV} &\sim&
\bigg(1-\frac{T}{T_c}\bigg)^\beta ~, \nonumber\\
\Sigma_0(0) &\sim&
\bigg(1-\frac{T}{T_c}\bigg)^\nu ~, \nonumber\\
f_\pi(T) &\sim&
\bigg(1-\frac{T}{T_c}\bigg)^{\beta'} ~,
\end{eqnarray}
where $T<T_c$.
The critical exponents $\beta$, $\nu$ and $\beta'$ are numerically
extracted by using the $\chi^2$ fitting.
The order parameters $\cal O$
($=\langle\overline\psi\psi\rangle_{1\rm GeV}, \Sigma_0(0), f_\pi(T)$)
are fitted by the linear functional form
$\ln {\cal O}(T) = A + \gamma \ln(1-T/T_c)$, and the fitting
parameters are $A$ and $\gamma$ ($=\beta,\nu,\beta'$) with $T_c=169$
MeV.
We have good fittings, which is shown in Fig.~\ref{fig:exponents}.
The result is
\begin{equation}
\label{exponents}
\beta = 0.171 ~,~~
\nu = 0.497 ~,~~
\beta' = 0.507 ~.
\end{equation}
These values are different from those in mean field theories.
If we consider mean field theories, we would have the relation
$\beta=\nu$ since $\Sigma \sim \langle\overline\psi\psi\rangle$.

\subsection{Zero temperature case}
\begin{figure}[htbp]
\begin{center}
\epsfxsize=\xsize
\ \epsfbox{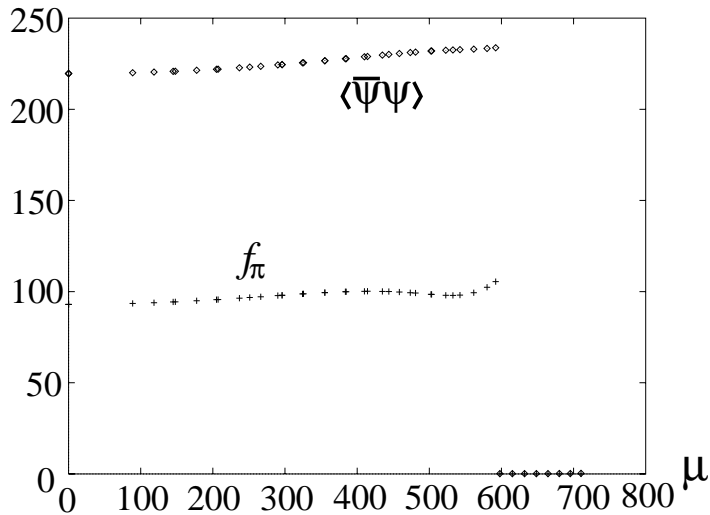}
\vspace{-5pt}
\caption[]{
The functional forms of the VEV and the pion decay constant for $\mu$
along $T=0$ line.
}
\label{fig:mu1}
\end{center}
\end{figure}

\begin{figure}[htbp]
\begin{center}
\epsfxsize=\xsize
\ \epsfbox{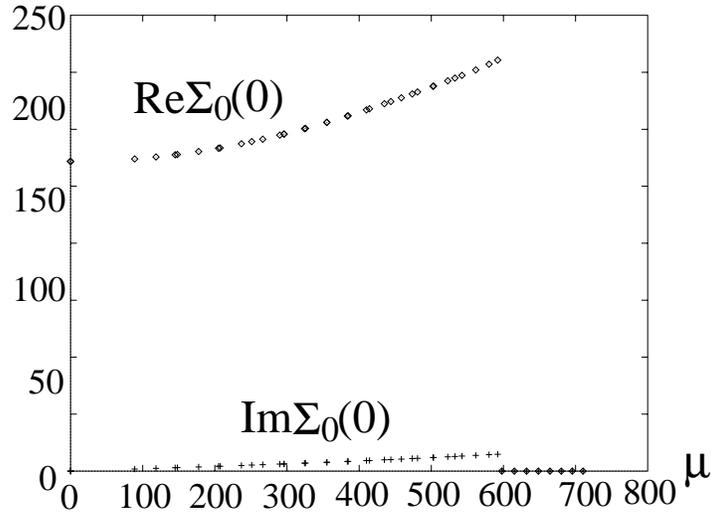}
\vspace{-5pt}
\caption[]{
The functional forms of the real and the imaginary parts of the mass
function for $\mu$ along $T=0$ line.
}
\label{fig:mu2}
\end{center}
\end{figure}

Next, we study the phase transition along the $T=0$ line.
As is seen from Eq.~(\ref{SDcompo}), the mass function has an
imaginary part for $\mu\neq0$.
The functional forms for $\mu$ are shown of the VEV, the pion decay
constant $f_\pi(\mu)$, the real and the imaginary parts of the mass
gap in Figs.~\ref{fig:mu1} and \ref{fig:mu2}.
The $SU(2)_L \times SU(2)_R$ chiral symmetry restores at
$\mu=\mu_c=598$ MeV.
We have a strong first order phase transition.
Here, we check that the infrared regularization parameter $t_F$ does
not affects the physical observables and the nature of the phase
transition.

Let us compare our result with those of other approaches.
There is no lattice simulation at so large chemical potential that we
can directly see the phase transition.
A phase transition is suggested by extrapolating the result of
lattice simulation around small $\mu$.\cite{mulattice}
On the other hand, there are many attempts using Schwinger-Dyson
equations\cite{BCS,Akiba,BCCGP} and NJL models\cite{AY,LKW}.
The previous attempts\cite{Akiba,BCCGP} in the ladder approximation
give a first order phase transition.
The NJL\cite{AY,LKW} models with the instanton effect also give the
same at $N_f=2,3$.
Therefore, our result confirms theirs.
Our advantage is that we have no parameter which modifies the physical
result.

\subsection{the phase diagram}
\begin{figure}[htbp]
\epsfxsize=7cm
\begin{center}
\ \epsfbox{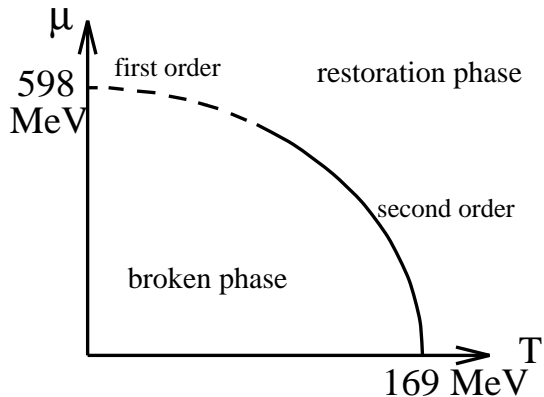}
\vspace{-5pt}
\caption[]{
The schematic view of the phase diagram from our result.
}
\label{fig:diagram}
\end{center}
\end{figure}

Finally, we study the phase diagram of the chiral symmetry
restoration.
Near the $\mu=0$ line we have second order phase transitions and near
the $T=0$ line we have first order ones.
In both cases the convergences of updating the mass function are rapid
well for solving the Schwinger-Dyson equation.
Unfortunately, near the phase transition line in the middle region,
the convergences are too bad to obtain solutions with a suitable
accuracy.
However, an natural guess will be that the order of phase transitions
continuously changes from  first order to second order, through weak
first order, in the middle region shown as in Fig.~\ref{fig:diagram}.
This type of diagram is also obtained in Refs.~\cite{BCCGP,AY}, which
is the same as that of two dimensional Gross-Neveu model in
Refs.~\cite{Wolff,IKM}.

\section{Summary and Discussion}
\label{sect:sam-dis}

In this paper we study the chiral symmetry restoration at finite
temperature and chemical potential in QCD.
We use the improved ladder approximation and the imaginary time
formalism.
The improved ladder approximation does not violate the chiral
symmetry, since the high energy behavior of the quark mass function is
consistent with the result of the operator product
expansion\cite{HigaMira} even at finite temperature and chemical
potential.
The phase transition point (or line) is determined by using the three
order parameters; i.e., the VEV
$\langle\overline\psi\psi\rangle_{1\rm GeV}$
renormalized at 1 GeV, the quark mass gap $\Sigma_0(0)$ and the pion
decay constant $f_\pi$.
In the improved ladder approximation the infrared regularization
parameter $t_F$ must be introduced as in Eq.~(\ref{g2}) in order to
regularize the running coupling.
We, however, observe that the physical quantities do not depend on the
parameter $t_F$.
Then, our results are obtained without any degrees of freedom other
than $\Lambda_{QCD}$ which is determined by putting $f_\pi = 93$ MeV
at $(T,\mu)=(0,0)$.

In the case of the vanishing chemical potential $\mu=0$ we have a
second order phase transition at $T_c=169$ MeV.
The critical exponents are extracted in Eq.~(\ref{exponents}).
This shows that the QCD in the improved ladder approximation is
different from mean field theories.
In the case of the vanishing temperature $T=0$ we have a strong first
order phase transition at $\mu = 598$ MeV.

In the $(T,\mu)=(0,0)$ limit the functional forms $A(p)=1$ and
$B(p)=0$ as in Eq.~(\ref{AB}) give the solution of the Schwinger-Dyson
equation in Landau gauge.
In this paper we put these forms for any $(T,\mu)$ for simplicity.
We should check the validity of this prescription.
In the middle region ($0<T<T_c$ and $0<\mu<\mu_c$) near the phase
transition line the calculation of the mass function is too hard for
the error to vanish in the iteration method.
It is necessary to obtain more efficient method for solving the
Schwinger-Dyson equation in this region.
After these problems being settled, the framework of the improved
ladder approximation becomes a more convenient tool to figure out the
nature of chiral symmetry, since it is easy to introduce fermions in
the chiral limit.

\begin{center}
\Large Acknowledgements
\end{center}
We would like to thank T. Hatsuda and Y. Kikukawa for valuable
discussions and comments.

\end{document}